# Effect of delay time on the generation of chaos in continuous systems

*Marek Berezowski*

Polish Academy of Sciences, Institute of Chemical Engineering

E-mail: marek.berezowski@polsl.pl

**Abstract**

The present study deals with a theoretical analysis of the effect of delay time of energy transport upon the generation of complex dynamics in continuous physical system. The importance of this time for the presence of quasi-periodicity and chaos in a reactor is demonstrated. The considerations are preceded by the analysis of one-dimensional mathematical model.

## 1. Introduction

The majority of physical objects is characterized, a.o., by the occurrence of delays of various types in the energy transport. If these delays are negligible compared to general changes in the process, then – as a rule – they are neglected in specific design calculations and in the mathematical analysis. It may be shown, however, that even small values of these delays may lead to qualitative changes in the nature of process dynamics [2]. One of the classical and instructive, in this respect, systems is a chemical reactor [1], [4] and [5]. In traditional computations it is assumed that the times transport through recirculation loop are equal to zero.

In the present work it has been demonstrated that in certain cases even the allowance for small value of delay time leads to very complex dynamic solutions, including chaos. The detailed theoretical considerations concerning the reactor. Non-zero time of mass transport through the recirculation loop has been assumed. The analysis has been based on the phenomena generated in logistic systems. In the case of non-zero delay time and lack of time derivations in reactor's model (the assumption justified only mathematically) the system of equations constitutes a logistic problem. The considerations concerning the reactor have been preceded by the theoretical analysis of a simple, one-dimensional mathematical model.

## 2. Transition from a logistic model to a continuous model

Let us take into account one of the frequently quoted logistic models [6]

$$x_{(k+1)} = ax_k(1 - x_k). \tag{1}$$

We assume that this simple model is connected with some physical process of definite inertia. So, this process may be written as a continuous one in the following way:

$$\sigma \frac{dx(\tau)}{d\tau} + x(\tau) = ax(\tau - \tau_d)[1 - x(\tau - \tau_d)], \tag{2}$$

where $\tau_d$ is the delay time. It may be clearly seen that for $\sigma=0$ the model (2) reduces identically to model (1), regardless of the value of $\tau_d$. As it is known, the logistic model generates, for certain values of parameter $a$, a complex dynamic solution, including chaos. Thus, there should exist some, arbitrarily small value of the parameter $\sigma$, below which the continuous model (2) also generates a complex dynamics solution, including chaos, too. In fact, assuming the parameters' values: $a=3.9, \tau_d=1$, the Feigenbaum diagram has been constructed. It results from the latter that in the solution of Eq. (2) chaos appears for the values of $\sigma<0.26$, approximately ( Fig. 1). This diagram indicates also that there exists a certain limiting value of $\sigma$, above which all oscillations disappear. This is clear, considering the fact that for large values of $\sigma$ the dynamics concentrates in inertia and not in delay. As it is well known, the one-dimensional model does not display oscillations. Equally interesting is the effect of delay time $\tau_d$ upon the generation of chaos. This is illustrated by Feigenbaum diagram (Fig. 2), where $a=3.9$ and $\sigma=0.2$ are assumed. Here, in turn, exists a limiting value of $\tau_d$, below which all oscillations disappear. The interpretation of this phenomenon is identical as previously. The third exemplary Feigenbaum diagram, connected with model (2), is depicted in Fig. 3, illustrating the effect of parameter $a$ on oscillations and chaos ($\sigma=0.2, \tau_d=1$). Comparing this graph with Feigenbaum diagram concerning the logistic model (1) it can be seen that the presence of parameter $\sigma$ shifts the interval of oscillations and chaos towards higher values of $a$. Having at our disposal the continuous model (2) we are able to determine analytically not only the values of parameters generating bifurcations (as may be done for the logistic model [6]) but also the value of the basic period of oscillations. A simple mathematical analysis leads, namely, to the statement is that the necessary and sufficient condition of generating oscillations in model (2) is

$$a > 2 + \sqrt{1+(\sigma\omega)^2}, \qquad (3)$$

the circular frequency $\omega$ being determined from the relationship

$$\tau_d\omega + \mathrm{arctg}(\sigma\omega) = \pi. \qquad (4)$$

It is worth nothing that $\omega$ does not depend on the value of $a$. For example, the temporal course of the variable $x$ for $a=3.9$, $\sigma=0.2$ and $\tau_d=0.65$ is shown in Fig. 4. The basic period of oscillations, determined from (4) equals $T=2\pi/\omega=1.69$. It follows from Fig. 2 that for the assumed values of parameters one has to do with the doubled period, $T^{(2)}=3.38$.

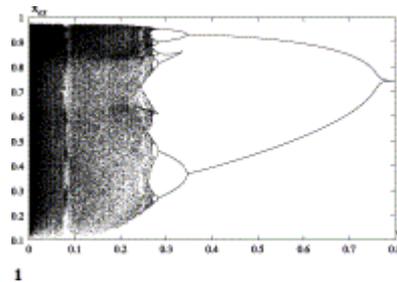

Fig. 1. Feigenbaum diagram of one-dimensional model. $a=3.9$, $\tau_d=1$.

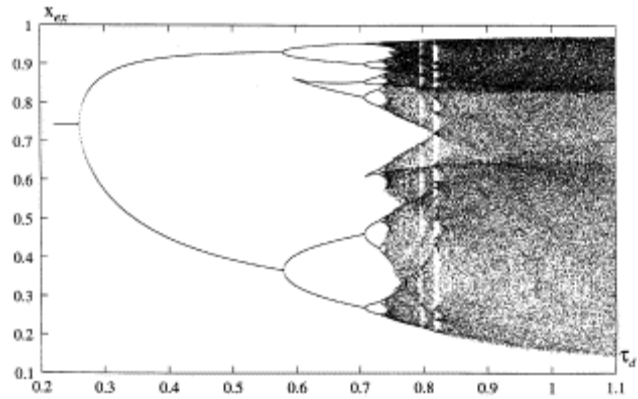

Fig. 2. Feigenbaum diagram of one-dimensional model. $a=3.9$, $\sigma=0.2$.

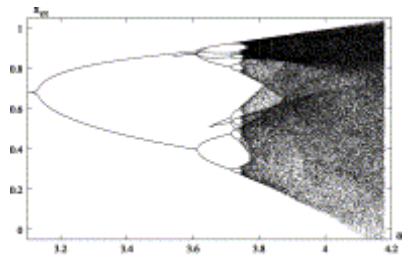

Fig. 3. Feigenbaum diagram of one-dimensional model. $\sigma = 0.2$, $\tau_d = 1$.

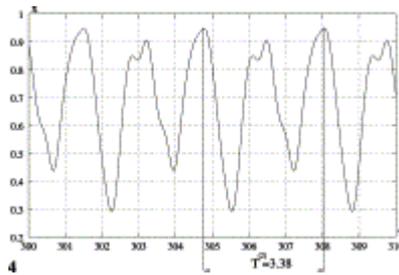

Fig. 4. One-dimensional-model. Temporal course. $a=3.9$, $\sigma = 0.2$, $\tau_d = 0.65$

The end of this part of the present study is devoted to the determination of the phase course of the chaotic solution of Eq. (2) (Fig. 5) as well as the Poincaré section of this solution (Fig. 6). This section is defined as extremes of the variable $x(\tau)$.

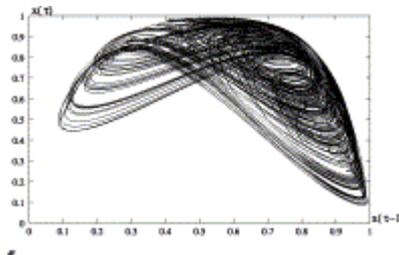

Fig. 5. One-dimensional-model. Strange attractor. $a=4$, $\sigma = 0.2$, $\tau_d = 1$.

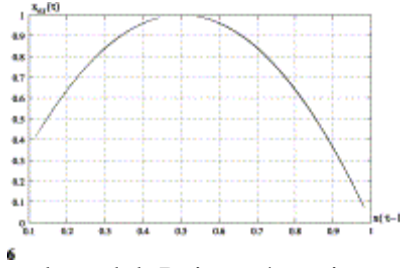

Fig. 6. One-dimensional-model. Poincaré section. $a=4$, $\sigma = 0.2$, $\tau_d = 1$.

## 3. Model of the reactor as extension of the logistic model

The mathematical model of the chemical reactor is represented by the following relationships:

Balance equations

$$\frac{\partial \alpha}{\partial \tau} + \frac{\partial \alpha}{\partial \xi} = \frac{1}{Pe_M}\frac{\partial^2 \alpha}{\partial \xi^2} + \phi, \qquad (5)$$

$$Le\frac{\partial \Theta}{\partial \tau} + \frac{\partial \Theta}{\partial \xi} = \frac{1}{Pe_H}\frac{\partial^2 \Theta}{\partial \xi^2} + \phi + \delta(\Theta_H - \Theta). \qquad (6)$$

Boundary conditions

for $\xi=0$

$$\frac{1}{Pe_M}\frac{d\alpha}{d\xi} = \alpha - f\alpha(1, \tau - \tau_d), \qquad (7)$$

$$\frac{1}{Pe_H}\frac{d\Theta}{d\xi} = \Theta - f\Theta(\tau - \tau_d), \qquad (8)$$

for $\xi=1$

$$\frac{d\alpha}{d\xi} = 0, \quad \frac{d\Theta}{d\xi} = 0, \qquad (9)$$

the $\phi$ function

$$\phi = (1-f)Da(1-\alpha)^n \exp\left(\gamma\frac{\beta\Theta}{1+\beta\Theta}\right). \qquad (10)$$

In the design calculations it is assumed, as a rule, that the delay time in mass transport through the recycle loop is just enough small compared with the residence time in the reactor that $\tau_d=0$ can practically be assumed. However, as it has been shown previously (as well as in the literature [2]) this time (even its small value) may be essential at the generation of complex dynamic solutions of the model, including chaos; viz., let us assume hypothetically that in the model , , , and the temporal derivatives are absent (this does not mean the steady state, but some mathematical simplification) and $\tau_d>0$. Then it can be written as

$$\frac{d\alpha_{(k+1)}}{d\xi} = \frac{1}{Pe_M} \frac{d^2\alpha_{(k+1)}}{d\xi^2} + \phi_{(k+1)}, \tag{11}$$

$$\frac{d\Theta_{(k+1)}}{d\xi} = \frac{1}{Pe_H} \frac{d^2\Theta}{d\xi^2} + \phi_{(k+1)} + \delta\left[\Theta_H - \Theta_{(k+1)}\right]. \tag{12}$$

Boundary conditions

for $\xi=0$

$$\frac{1}{Pe_M} \frac{d\alpha_{(k+1)}}{d\xi} = \alpha_{(k+1)} - f\alpha_k, \tag{13}$$

$$\frac{1}{Pe_H} \frac{d\Theta_{(k+1)}}{d\xi} = \Theta_{(k+1)} - f\Theta_k, \tag{14}$$

for $\xi=1$

$$\frac{d\alpha_{(k+1)}}{d\xi} = 0, \quad \frac{d\Theta_{(k+1)}}{d\xi} = 0. \tag{15}$$

The system (11)-(15) and constitutes a typical two-dimensional logistic model, which may generate chaos. Hence, on the ground of previous considerations it may be stated that there is a chance that the chaotic solutions, in the case of presence of temporal derivatives, will hold in some range of the values of the parameters of reactor's model, e.g., up to a certain value of Lewis number *Le*. In order to prove it, the simulation computations of the model (5)-(9) and have been performed aimed at the determination of Feigenbaum diagrams. So in Fig. 7a diagram is depicted. It results from the latter that chaos appears already at a relatively small value of delay time in the recycle loop, $\tau_d>0.06$ (although there is no chaos for $\tau_d=0$!). This value certainly would be disregarded in design calculations resulting in the obtaining of completely erroneous dynamic results; viz., for $\tau_d=0$ the periodic oscillations are generated in the reactor (Fig. 7). Fig. 8 presents the three-dimensional Feigenbaum diagram illustrating the effect of *Le* number upon the generation of chaos. For the assumed values of the parameters this effect appears in the range $1<Le<1.0125$ (this means that it practically appears in the homogeneous reactor). On the other hand, for somewhat higher values of Lewis number ($Le>1.0465$) an interesting dynamic solution appears, which is presented in Fig. 9. It turns out that this is a quasi-periodic solution [7]. The evidence is, a.o., visible elliptic contours in the section $\tau_d$=const. These contours develop from Hopf bifurcation points A–F. Based on Fig. 7 the phase trajectories of the reactor's model for $\tau_d=0.2$ (Fig. 10) have been determined. These trajectories confirm the presence of chaos in the mentioned range of parameters' value; viz., one has here to do with the so-called strange attractor. Analogous trajectories have also been determined on the ground of Fig. 9, at the assumption $\tau_d=0.05$ and $\tau_d=0.2$. This time, however, the attractors are markedly ordered Fig. 11 and Fig. 12, which bears witness to the quasi-periodicity of the temporal cause. In order to confirm the foregoing results, the so-called Poincaré sections have been executed. They are defined as the extremes of temperature at reactor's outlet, $\Theta_{ex}(1)$ [3]. A typical Poincaré section of chaotic solution is shown in Fig. 13. Fig. 14 presents the enlargement of a fragment of this section, indicating multilayer structure of the graph. Analogous sections have also been executed for the quasi-periodic state Fig. 15 and Fig. 16. In Fig. 15 four closed contours were visible, whereas in Fig. 16 – two ones. The number of these contours follows directly from the number of bifurcation points

HB; viz., Fig. 15 refers to the left-hand side of Feigenbaum diagram (Fig. 9), 4 HB, whereas Fig. 16 concerns the right-hand side (2 HB). Four contours from Fig. 15 testify that the phase trajectory moves along four toruses. On the other hand, two contours from Fig. 16 bear witness to the fact that the phase trajectory moves along two toruses, passing from – time to time – from one torus to the second one.

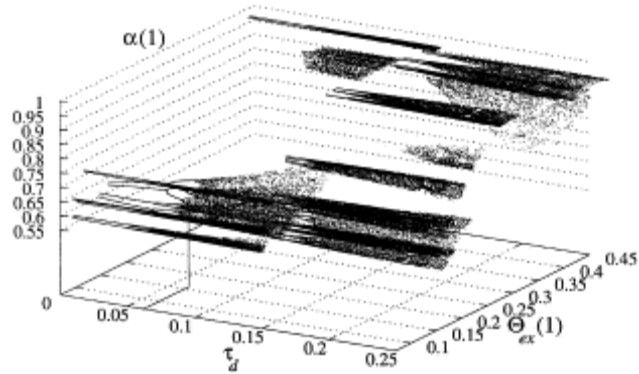

Fig. 7. Three-dimensional Feigenbaum diagram of the reactor with recycle. $Da$=0.15, $\gamma$=15, $\beta$=2, $\delta$=3, $n$=1.5, $Pe_M = 100$, $Pe_H = 100$, $Le = 1$, $f = 0.7$, $\Theta_H$=−0.0268.

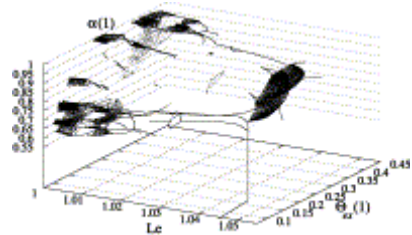

Fig. 8. Three-dimensional Feigenbaum diagram of the reactor with recycle. $Da$=0.15, $\gamma$=15, $\beta$=2, $\delta$=3, $n$=1.5, $Pe_M = 100$, $Pe_H = 100$, $\tau_d = 0.2$, $f = 0.7$, $\Theta_H$=−0.0268.

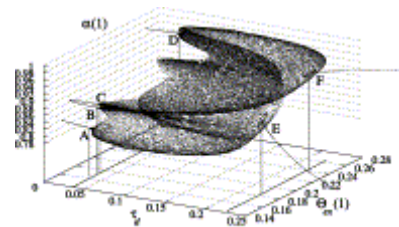

Fig. 9. Three-dimensional Feigenbaum diagram of the reactor with recycle. $Da$=0.15, $\gamma$=15, $\beta$=2, $\delta$=3, $n$=1.5, $Pe_M = 100$, $Pe_H = 100$, $Le = 1.05$, $f = 0.7$, $\Theta_H$=−0.0268.

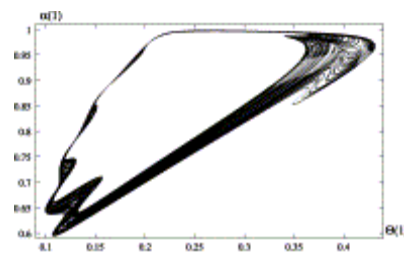

Fig. 10. Reactor with recycle. Strange attractor. $Da$=0.15, $\gamma$=15, $\beta$=2, $\delta$=3, $n$=1.5, $Pe_M = 100$, $Pe_H = 100$, $Le = 1$, $\tau_d = 0.2$, $f = 0.7$, $\Theta_H$=−0.0268.

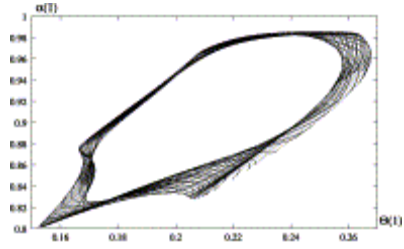

Fig. 11. Reactor with recycle. Attractor of quasi-periodic course. *Da*=0.15,
γ=15,β=2,δ=3,n=1.5, $Pe_M = 100$, $Pe_H = 100$, $Le = 1.05$, $\tau_d = 0.05$, $f = 0.7$, $\Theta_H$=−0.0268.

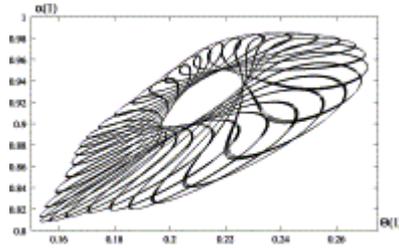

Fig. 12. Reactor with recycle. Attractor of quasi-periodic course. *Da*=0.15,
γ=15,β=2,δ=3,n=1.5, $Pe_M = 100$, $Pe_H = 100$, $Le = 1.05$, $\tau_d = 0.2$, $f = 0.7$, $\Theta_H$=−0.0268.

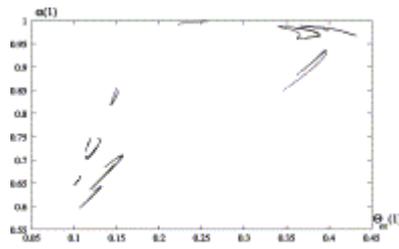

Fig. 13. Reactor with recycle. Poincaré section. Chaos. *Da*=0.15, γ=15,β=2,δ=3,n=1.5,
$Pe_M = 100$, $Pe_H = 100$, $Le = 1$, $\tau_d = 0.2$, $f = 0.7$, $\Theta_H$=−0.0268.

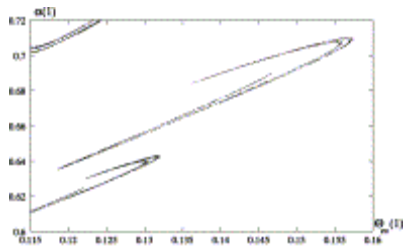

Fig. 14. Detail of Poincaré section from Fig. 13.

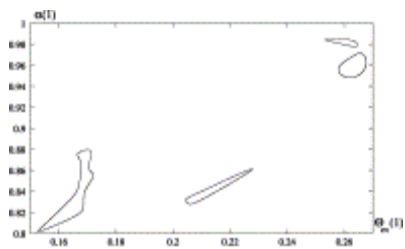

Fig. 15. Reactor with recycle. Poincaré section. Quasi-period.
$Da = 0.15$, $\gamma = 15$, $\beta = 2$, $\delta = 3$, $n = 1.5$, $Pe_M = 100$, $Pe_H = 100$, $Le = 1.05$, $\tau_d = 0.05$, $f = 0.7$
, $\Theta_H$=−0.0268.

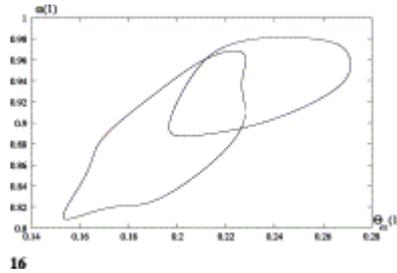

Fig. 16. Reactor with recycle. Poincaré section. Quasi-period. *Da*=0.15, $\gamma=15$, $\beta=2$, $\delta=3$, $n=1.5$, $Pe_M=100$, $Pe_H=100$, $Le=1.05$, $\tau_d=0.2$, $f=0.7$, $\Theta_H$=−0.0268.

Finally, it should be mentioned that in the case of a homogeneous reactor (*Le*=1, [1] and [4]) without dispersion ($Pe_M=\infty$, $Pe_H=\infty$), the complete dynamic model , , and (condition (9) does not hold here) constitutes the logistic model and the quality of its solutions does not depend on the delay time $\tau_d$. In this case $\tau_d$ affects solely the value of period of oscillations.

## *4. Concluding remarks*

This work presents the effect of the delay time of reactor upon the generation of quasi-periodicity and chaos in the system. It has been shown that – in some cases – even an insignificant value of this time may lead to complex dynamic solutions of the object under study. The result as the above is of essential practical importance, taking into account that in the majority of design cases $\tau_d$=0 is assumed. This assumption yields as a result a completely different (i.e., erroneous) temporal solutions. The considerations concerning reactor are preceded by the theoretical analysis of one-dimensional model. The transition from the logistic case to the continuous one has been analyzed.